\documentclass[a4paper,12pt]{article}
\newif\ifpdf            
\ifx\pdfoutput\undefined \pdffalse \else \pdftrue \fi
\ifpdf 
\usepackage[pdftex]{graphicx}
\pdfcompresslevel9 \DeclareGraphicsExtensions{.jpg,.pdf,.png,.mps}
\else 
\usepackage[dvips]{graphicx}
\DeclareGraphicsExtensions{.eps,.ps} \fi
\usepackage{bm}

\newcommand{\beq}{\begin{equation}}
\newcommand{\eeq}{\end{equation}}

\newcommand{\lzero}{\ensuremath{\underline{\underline{\ell}}_0}}
\newcommand{\ltheta}{\ensuremath{\underline{\underline{\ell}}_{\theta}}}
\def\matr#1{\ensuremath{\underline{\underline{{\bm{#1}}}}}}

\begin{document}

\title{Tube Model for the Elasticity of Entangled Nematic Rubbers}

\author{S.~Kutter and
E.M.~Terentjev\footnote{Electronic mail: {\it emt1000@cam.ac.uk}} \\
Cavendish Laboratory, University of Cambridge \\ Madingley Road,
Cambridge CB3 0HE, U.K.}

\date{\today}
\maketitle

\abstract{Dense rubbery networks are highly entangled polymer
systems, with significant topological restrictions for the
mobility of neighbouring chains and crosslinks preventing the
reptation constraint release. In a mean field approach,
entanglements are treated within the famous reptation approach,
since they effectively confine each individual chain in a
tube-like geometry. We apply the classical ideas of reptation
dynamics to calculate the effective rubber-elastic free energy of
anisotropic networks, nematic liquid crystal elastomers, and
present the first theory of entanglements for such a material.}

\vspace{0.2cm}
\noindent {PACS numbers:} \\
      {61.41.+e} \ {Polymers, elastomers, and plastics}   \\
      {61.20.Vx} \ {Polymer liquid crystals} \\
        {62.20.Dc} \ {Elasticity, elastic constants}

\newpage

\section{Introduction}

Rubbery polymer networks are complex randomly disordered amorphous
systems. The simplest theoretical models consider them as being
made of ``phantom chains'', where each chain is thought to be a
three-dimensional random walk in space. To form a network, the
chains are crosslinked to each other at their end points, but do
not interact otherwise, in particular they are able to fluctuate
freely between crosslinks. This has an unphysical consequence that
the strands can pass through each other. If one tries to avoid
this assumption, the theory is confronted with the intractable
complexity of entanglements and their topological constraints. The
mean field treatment of entangled polymer melts and semi-dilute
solutions is the classical reptation theory
\cite{gennes79,doiedw86} going back to the early seventies, which
has been a spectacular success in describing a large variety of
different physical effects. However, the parallel description of
crosslinked rubbery networks has been much less successful. First
of all, there is a significant difference in entanglement
topology: in a melt the confining chain has to be long enough to
form a topological knot around a chosen polymer; even then the
constraint is only dynamical and can be released by a reptation
diffusion along the chain path. In a crosslinked network, any loop
around a chosen strand becomes an entanglement, which could be
mobile but cannot be released altogether. A number of other
complexities arise from the coupling between imposed deformations
and chain anisotropy, the stress-optical effects
\cite{jarry-monnerie79,deloche-samulski81,doi-pearson89} and
nematic interactions between chain segments
\cite{bladon-warner93,abramchuk-nyrkova89}.

In addition, the polymer network can be spontaneously anisotropic,
forming a liquid crystalline elastomer (LCE). This area has
attracted a significant experimental and theoretical interest in
recent years. In nematic LCE, the strands preferably orient
themselves along one direction, forming a nematic liquid crystal
order. The response to an external deformation is now of a much
richer nature, with antisymmetric stress and internal torques
depending on the relative angle of the director to the axis of
deformation \cite{warter96}. Liquid crystalline elastomers combine
remarkable properties of both its components, liquid crystals and
rubbers, but also show physical properties that place them in a
separate category from any other material. Several new physical
phenomena have been discovered in LCE: (a) spontaneous, reversible
shape changes of up to 400\% on temperature change; (b) ``soft
elasticity'' -- mechanical deformation, involving modifications of
internal nematic microstructure, without (or with very low)
stress; (c) mechanical and electric instabilities involving
director reorientation, in special cases discontinuous jumps; (d)
solid phase nematohydrodynamics and unusual rheology, leading to
anomalous dissipation and acoustic effects. Recent reviews
\cite{brafin98,ter99} describe the current state of affairs in
this field. Our challenge in this paper is to bring the
microscopic theoretical description of nematic rubbers on the same
level as in the classical isotropic rubbers, in particular, to
account for chain entanglements.

An early model of elastic response of entangled rubbers was
developed by Edwards \cite{edward77}: in tradition with the melt
theory, it assumed that the presence of neighbouring strands in a
dense network effectively confines a particular polymer strand to
a tube, whose axis defines the primitive path. Within this tube,
the polymer is free to explore all possible configurations,
performing random excursions, parallel and perpendicular to the
axis of the tube. One can show that on deformation of the sample
the length of the primitive path increases. Since the arc length
of the polymer is constant, the amount of chain available for
perpendicular excursions is reduced, leading to a reduction in
entropy and an increase in rubber-elastic free energy. A number of
further attempts have been made to derive a self-consistent theory
of entangled rubber elasticity. Of this list, the most significant
are the scaling ``localisation model'' of Gaylord and Douglas
\cite{gaydou90}, the ``slip-link model'' of Ball, Doi, Edwards and
Warner (BDEW) \cite{baldoi81}, accounting for entanglements as
local mobile confinement sites linking two interwound strands, and
the ``hoop model'' by Higgs and Ball (HB) \cite{higbal89}, who
assumed that entanglements localise certain chain segments. In an
article developing the reptation theory of rubber elasticity for
classical isotropic networks \cite{ouriso}, we give a more
detailed overview of these and other theories.

In our current work, we extend the tube model to treat the
elasticity of anisotropic networks of liquid crystalline polymers.
To our knowledge, this is the first time that a reptation model
has been applied to treat the effects both of the entanglements
and of the anisotropic nature of the nematic network. The tube
model provides a more accurate microscopic description in the
sense that it keeps track of the allocation of chain segments and
their excursions between the points of entanglement. The next
Section briefly reviews the ideal phantom-network approach to the
elasticity of nematic rubber and introduces the tube model and its
properties, giving the full expression for nematic rubber-elastic
free energy. Section~3 contains the discussion of the model and
its results, including the linear-response limit. We conclude by
comparing the results of the present theory with those of the
ideal phantom network and analyse which physical properties of LCE
seem to be most sensitive to the effect of chain entanglements.

\section{Nematic elastomer network}

Before developing our model for macroscopic elasticity of densely
entangled rubber, we briefly review the well-known results of the
phantom chain network theory, which provides the basics to most
other theoretical models.

\subsection*{Phantom chain approximation}

Assuming that a single polymer performs a free random walk in
three dimensions, one finds that the end-to-end distance ${\bm
R}_0$ obeys  a Gaussian distribution in the long chain limit. This
result goes back far in history: one can review its derivation and
consequences in the classical text on this subject
\cite{doiedw86}. The distribution of $\bm{R}_0$ is given by
\begin{eqnarray}
    P_0({\bm R}_0)= \left(\frac{3}{2\pi N
    b^2}\right)^{3/2}\exp\left(-\frac{3}{2 N b^2}{\bm R}_0^2\right),
     \label{gauss}
\end{eqnarray}
where $b$ is the monomer step length and $N$ the number of steps
of one chain trajectory.

In a nematic polymer, irrespective of the particular mesogenic
mechanism, the monomer steps acquire a preferred orientation along
the director $\bm{n}_0$. Accordingly, the end-to-end distance
distribution function of a strand becomes anisotropic as well:
 \begin{eqnarray}
P(\bm{R_0})&=& \left(\frac{3}{2\pi N
b}\right)^{3/2}(\det\lzero)^{-1/2}      \label{aniso-distr} \\
&& \times    \exp\left(-\frac{3}{2 L}
\bm{R}_0^{\mathsf{T}}\cdot\lzero^{-1}\cdot\bm{R}_0\right),
\nonumber
 \end{eqnarray}
where $Nb=L$ is the contour length of the chain, and the matrix
\lzero takes account of the anisotropy:
$$
    (\lzero)_{ij}=
    l_0^{\perp}\delta_{ij}+(l_0^{\parallel}-l_0^{\perp})n_{0i}n_{0j}.
$$
This matrix of anisotropic chain steps is directly measurable from
the average chain shape, given by $\langle R_i R_j \rangle =
\frac{1}{3}(\lzero)_{ij} L$.  The principal values of this
effective step-length matrix, $l_0^{\perp}$ and $l_0^{\parallel}$,
reflect the spontaneous nematic order in the material. In the
isotropic phase, e.g., above the nematic transition temperature
$T_{\rm ni}$, $l_0^{\parallel}= l_0^{\perp}= b$ and one trivially
recovers the isotropic Gaussian distribution (\ref{gauss}). The
difference $l_0^{\parallel}- l_0^{\perp}$ is proportional to the
nematic order parameter $Q$. However, the explicit form of this
dependence is different in different models of nematic polymers.
In a most simple case of freely jointed chain of rods of length
$b$ one obtains $l_0^{\parallel}=b(1+2Q), \ l_0^{\perp}=b(1-Q)$,
while in the hairpin regime of semiflexible main-chain nematic
polymer the anisotropy could become very large: $l_0^{\parallel}
\propto \exp [3/(1-Q)]$, cf. \cite{wang-warner86}. The power of
the ideal theory of nematic rubber elasticity \cite{warter96} is
in that it is independent of such model considerations and only
uses a single model parameter -- the ratio
$r=l_0^{\parallel}/l_0^{\perp}$, or equivalently, $r=\langle
R_\parallel^2 \rangle / \langle R_\perp^2 \rangle$ for the
principal values of the gyration radius. We shall see below that
this attractive feature is reproduced in the theory of entangled
nematic networks.

The entropic free energy of such an anisotropic random walk is
given by the logarithm of the number of conformations with the
fixed $\bm{R}_0$ and has the form $$\beta F= -\ln P({\bm
R}_0)=\frac{3}{2L} \bm{R}_0^{\mathsf{T}} \cdot\lzero^{-1}
\cdot\bm{R}_0 + \mathrm{const},$$ where $\beta=1/k_B T$ is the
inverse Boltzmann temperature. At formation of the network, i.e.
at crosslinking, the polymer melt is assumed to obey the
anisotropic Gaussian distribution (\ref{aniso-distr}), which is
then permanently frozen in the network topology. In the phantom
network approximation, the lateral restrictions on the chain
thermal motion are neglected and different network strands
interact only at the crosslinking points.

One then assumes that the junction points deform affinely with
respect to their initial positions $\bm{R}_0$ by the macroscopic
deformation \matr{\lambda}, hence we can write
$\bm{R}=\matr{\lambda} \bm{R}_0$. Therefore, the deformation
$\matr{\lambda}$ alters the free energy of each strand. The change
of free energy per chain of the whole network can be calculated by
the usual quenched averaging:
 \beq
   \beta F =-\langle\ln P(\matr{\lambda} \bm{R}_0)\rangle_{P(\bm{R}_0)}
   =\frac{1}{2} \mathrm{Tr}(\lzero\cdot\matr{\lambda}^{\mathsf{T}}\cdot
   \ltheta^{-1}\cdot\matr{\lambda}), \label{fgauss}
 \eeq
where we have dropped an irrelevant constant and found a new
expression for the chain step-length matrix after deformation:
$$
    (\ltheta)_{ij}=
    l^{\perp}\delta_{ij}+(l^{\parallel}-l^{\perp})n_{i}n_{j},
$$
with the rotated director, $(\bm{n}\cdot \bm{n}_0)=\cos \theta$,
and possibly changed principal values $l^{\parallel}$ and
$l^{\perp}$. The overall elastic free energy density, in the first
approximation, is simply (\ref{fgauss}) multiplied by the number
of elastically active network strands in the system $n_{\rm ch}$
per unit volume, which is proportional to the crosslinking
density:
 \beq
F_{\rm el}=\frac{1}{2}\mu \
    \mathrm{Tr}(\lzero\cdot\matr{\lambda}^{\mathsf{T}}\cdot\ltheta^{-1}\cdot\matr{\lambda}),
    \label{ftrace}
\eeq with the rubber modulus $\mu=n_{\rm ch}k_B T$, cf.
\cite{warter96} for detail.

The phantom-network model of rubber elasticity is a popular first
approximation. There are several reasons for its overall success
in spite of obvious oversimplifications. The crosslinking points
connect the ends of different strands together and thus reduce
local fluctuations -- and, therefore, alter the single chain
statistics. However, in spite of an apparent complexity of this
problem, it has been shown \cite{wargel88} that this effect merely
introduces a trivial multiplicative factor of the form $1-2/\phi$,
where $\phi$ is the junction point functionality. Secondly, one
can assume that the deformation preserves the volume, since the
bulk (compression) modulus is by a factor of at least $10^4$
greater than the shear modulus, which is proportional to $\mu$;
this implies the constraint $\det\matr{\lambda}=1$. Thirdly, the
quenched average in equation (\ref{fgauss}) does not average over
chains of different arc lengths, but the fact that the result is
independent of arc length, generalises the result to apply for
chains of arbitrary length, or even for a polydisperse ensemble of
chains. In the particular case of nematic LCE, this simple model
provides a rich crop of theoretical predictions described in
greater detail in quoted review articles.

\subsection*{The tube model}

\begin{figure}
\centering \resizebox{0.47\textwidth}{!}{\includegraphics{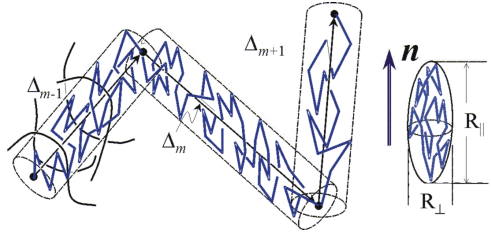}}
\caption{A nematic polymer strand is surrounded by neighbouring
chains, which effectively confine the strand to a tube. The tube
segment $m$ contains $s_m$ monomer steps, where the index $m$ runs
from 1 to $M$. Since the direction of chain steps is, on average,
uniaxial (as illustrated by the drawing of an ellipsoidal shape of
a free chain gyration on the right), the number of steps in each
tube segment depends on the orientation of $\bm{\Delta}_m$ with
respect to the nematic director $\bm{n}$. } \label{pic}
\end{figure}

Following the original ideas of Edwards \cite{edward77}, we assume
that one particular network strand is limited in its lateral
fluctuations by the presence of neighbouring chains. Therefore
each segment of a given strand only explores configurations in a
limited volume, which is much smaller than in the random coil
state. Hence, the whole strand fluctuates around a mean path,
which we call the primitive path. Effectively, the chain is
confined to exercise its thermal motion only within a tube around
the primitive path due to the presence of neighbouring chains.
This primitive path itself can be considered as a random walk with
an associated typical step length, which is much bigger than the
monomer step length \cite{viledw}. The step length of the
primitive path divides the tube into tube segments, as sketched in
Fig.~\ref{pic}, and therefore determines the number of tube
segments $M$ along one polymer strand.

Note that all chains are in constant motion, altering the local
constraints they impose on each other. Hence, the tube is a gross
simplification of real situation. However, one expects this to be
an even better approximation in rubber than in a corresponding
melt (where the success of reptation theory is undeniable),
because the restriction on chain reptation diffusion in a
crosslinked network eliminates the possibility of constraint
release.

To handle the tube constraint mathematically, we assume that the
chain segments are subjected to a quadratic potential, restricting
their motion transversely to the primitive path. Along one polymer strand
consisting of $N$ mo\-no\-mers of effective step length $b$, there are
$M$ tube segments, each containing $s_m,\ m=1,\dots,M$ monomer
steps. We infer the obvious condition
 \beq
    \sum_{m=1}^{M} s_m=N.\label{segmentsum}
 \eeq
In effect, one has two random walks: the topologically fixed
primitive path and the polymer chain restricted to move around it
-- both having the same end-to-end vector $\bm{R}_0$, between the
connected crosslinking points.

Each tube segment $m$ can be described by the span vector
$\bm{\Delta}_m$, joining the equilibrium positions of the strand
monomers at the two ends of each tube segments. The number of tube
segments $M$ (or, equivalently, the number of chain entanglements,
$M-1$) is a free parameter of the theory, ultimately determined by
the length of each polymer strand and the average ``entanglement
density''.

Since the primitive path is a topologically frozen characteristic
of each network strand, we shall assume that all primitive path
spans $\bm{\Delta}_m$ deform affinely with the macroscopic strain:
$\bm{\Delta}_m'=\matr{\lambda} \bm{\Delta}_m$. This is the central
point in the model: the rubber elastic response will arise due to
the change in the number of polymer configurations in a distorted
primitive path. To evaluate the number of conformations, we look
separately at chain excursions parallel and perpendicular to the
tube axis, within each span $\bm{\Delta}_m$. Effectively, this
amounts to introducing a new coordinate system for each tube
segment, with one preferred axis along $\bm{\Delta}_m$. The
constraints exerted by the other chains only constrain the
considered polymer in its lateral motion. Hence, we recover the
behaviour of a one-dimensional random walk in the direction of
$\bm{\Delta}_m$, giving rise to Gaussian statistics in the long
chain limit. Note that only one third of the steps $s_m$ in this
tube is involved in the longitudinal excursions. We therefore
obtain for the number of longitudinal excursions in a tube segment
$m$, cf. Fig.~\ref{pic},
 \beq
    W_m^{\rm (L)}\propto
    \frac{1}{\sqrt{s_m/3}}
    \exp\left(-\frac{1}{2 b (s_m/3)}
    \bm{\Delta}_m^{\mathsf{T}}\cdot\lzero^{-1}\cdot\bm{\Delta}_m
    \right).
    \label{parallel}
 \eeq
The spontaneous anisotropy of nematic polymer chain is reflected
in (\ref{parallel}) by accounting for the difference in number of
chain conformations in a given tube segment, depending on its
orientation with respect to the local nematic director (the
principal axis of step-length matrix $\lzero$).

To determine the number of transverse excursions, we introduce the
Green's function for the steps made by the chain in the plane
perpendicular to the local tube axis $\bm{\Delta}_m$. In effect,
we consider a two-dimensional random walk, with a total number of
steps $(2s_m/3)$, in a centrosymmetric quadratic potential. For
each of these two perpendicular coordinates, the Green's function
satisfies the following modified diffusion equation (see e.g.
\cite{doiedw86}, and its extension for the uniaxial nematic case
in \cite{wargel88}). The argument that follows, about transverse
chain movements in a confining potential, has a very simple
conclusion -- that the conformational effects are irrelevant for
calculation of rubber elasticity and only important factor is the
number of chain steps attributed to this degree of freedom.
Equally, the anisotropic (nematic) nature of chain random walk
does not contribute to the entropy of strongly confined transverse
excursion. Although the full anisotropic treatment is possible,
here we shall use a much shorter and transparent version of the
isotropic chain confined in the tube; its Green's function
satisfies the differential equation for each of the two
coordinates:
 \beq
    \left(\frac{\partial}{\partial s}-
    \frac{1}{2}b^2\frac{\partial^2}{\partial x_f^2}
    +\frac{1}{2}q_0^2x_f^2\right)G(x_i,x_f;s)=\delta(x_f-x_i)\delta(s),
\label{greeneq}
 \eeq
where the $x_i$ and $x_f$ are the initial and final coordinates of
the random walk with respect to the tube axis  and $q_0$
determines the strength of the confining potential. The equation
(\ref{greeneq}) is very common in the physics of polymers and its
exact solution is known. However, we only need to consider a
particular limit $q_0 b s_m \gg 1$ of this solution, which is the
case of dense entanglements (resulting in a strong confining
potential) and/or of a large number $s_m$ of monomers confined in
the tube segment. Outside this limit, that is, when the tube
diameter is the same order as the arc length of the confined
chain, the whole concept of chain entanglements becomes
irrelevant. In the strongly confined limit the solution has a
particularly simple form \cite{edward77}:
 \beq
    G(x_i,x_f;s)\propto
    \exp\left(-\frac{q_0}{2 b}(x_i^2+x_f^2)-\frac{1}{6}q_0 b s_m\right).
 \eeq
Remembering that there are two coordinates describing the
transverse excursions, we obtain for the two-di\-men\-sio\-nal Green's
function of the tube segment $m$:
 \beq
    G_m(\bm{r}_i,\bm{r}_f;s)\propto
    \exp\left(-\frac{1}{3}q_0 b s_m\right)
    \exp\left(-\frac{q_0}{2 b}(\bm{r}_i^2+\bm{r}_f^2)\right),
    \label{2d-green}
 \eeq
where $\bm{r}_i$ and $\bm{r}_f$ are the initial and final
transverse two-dimensional coordinates.

The total number of transverse excursions is proportional to the
integrated Green's function: $$ W_m^{\rm(T)}\propto\int d\bm{r}_i
\int d\bm{r}_f G_m(\bm{r}_i,\bm{r}_f;s).$$ Since the Green's
function (\ref{2d-green}) does not couple the initial or final
coordinates to the number of segments $s_m$, this integration will
only produce a constant normalisation factor which can be
discarded. Exactly the same conclusion is reached if the modified
diffusion equation for anisotropic chain is considered.

Gathering the expressions for statistical weights of parallel and
perpendicular excursions (and returning to the fully anisotropic
description), one obtains the total number of configurations of a
polymer segment consisting of $s_m$ monomers in a tube segment of
span $\bm{\Delta}_m$:
 \begin{eqnarray}
    W_m &=& W_m^{\rm (L)} W_m^{\rm (T)} \label{fullW} \\
    &\propto &
    \frac{1}{\sqrt{s_m}}
    \exp\left(- \frac{1}{3}q_0 bs_m -\frac{1}{2 b (s_m/3)}
    \bm{\Delta}_m^{\mathsf{T}}\cdot\lzero^{-1}\cdot\bm{\Delta}_m
    \right).\nonumber
 \end{eqnarray}
Therefore, we find for the full number of configurations of the
whole strand
 \beq
 W= \int\limits_0^N ds_1 \cdots \int\limits_0^N ds_M
    \left(\prod_{m=1}^{M} W_m \right)
    \delta\left(\sum_{m=1}^{M} s_m-N\right),\label{no_config}
 \eeq
where we have implemented the polymer contour length constraint
(\ref{segmentsum}). The statistical summation in (\ref{no_config})
takes into account the reptation motion of the polymer between its
two crosslinked ends, by which the number of segments, $s_m$,
constrained within each tube segment can be changed and, thus,
equilibrates for a given conformation of primitive path.

Rewriting the delta-function as $\delta(x)=\frac{1}{2\pi} \int dk
\ e^{{\rm i} kx}$, we proceed by finding the saddle points $s_m^*$
which make the exponent of statistical sum (\ref{no_config})
stationary. It can be verified that the normalisation factors
$1/\sqrt{s_m}$ contribute only as a small correction to the saddle
points
 \beq
s_m^* \approx \left( \frac{3 \ \bm{\Delta}_m^{\mathsf{T}}\cdot
\lzero^{-1}\cdot \bm{\Delta}_m }{2b (\frac{1}{3}q_0b + {\rm
i}k)}\right)^{1/2}.\label{saddle-sm}
 \eeq
The integral in (\ref{no_config}) is consequently approximated by
the steepest descent method. We repeat the same procedure for the
integration of the single auxiliary variable $k$, responsible for
the conservation of the polymer arc length. The saddle point value
$k^*$, inserted into (\ref{saddle-sm}), gives
$$\overline{s}_m =\frac{N |\lzero ^{-1/2}\bm{\Delta}_m |}{\sum_{i=m}^{M}
 |\lzero ^{-1/2}\bm{\Delta}_m |} ,$$
which is the equilibrium number of steps the nematic polymer makes
in a tube segment characterised by the axis vector
$\bm{\Delta}_m$. By completing the saddle-point integration, we
finally obtain the total number of configurations of one strand,
confined within a tube whose primitive path is described by the
set of vectors $\{ \bm{\Delta}_m \}$. The statistical weight $W$
associated with this state is proportional to the probability
distribution:
\begin{eqnarray}
    \lefteqn{W(\bm{\Delta}_1,\dots,\bm{\Delta}_M) \propto
P(\{\bm{\Delta}_m\}) \label{weight} }\\
&\propto&  \frac{\exp \left(-\frac{3}{2bN}
\left(\sum_{m=1}^{M}|\lzero^{-1/2}\bm{\Delta}_m |\right)^2
-\frac{1}{3}q_0 b N \right)} {\left(\sum_{i=m}^{M}
|\lzero^{-1/2}\bm{\Delta}_m |\right)^{M-1}}. \nonumber
\end{eqnarray}
The scalar $|\lzero^{-1/2} \bm{\Delta}_m |$ reflects the length of
the $m$-th step of the primitive path, modified by its projection
on the uniaxial matrix of chain step-lengths. This expression is a
result parallel to the ideal Gaussian $P(\bm{R}_0)$ in equation
(\ref{aniso-distr}) for a unentangled chain. Note that the chain
end-to-end distance $\bm{R}_0$ is also the end-to-end distance of
the primitive path random walk: $\sum_{m=1}^M \bm{\Delta}_m
=\bm{R}_0$.

\subsection*{Free energy of deformations}

From the equation (\ref{weight}) we obtain the formal expression
for free energy of a chain confined to a tube with the primitive
path conformation $\{\bm{\Delta}_m\}$, $\beta F=-\ln W$, or
 \begin{eqnarray}
    \beta F &=& \frac{3}{2b N}
\left(\sum_{m=1}^M |\lzero^{-1/2}\bm{\Delta}_m | \right)^2
  \label{free0}\\
&& + (M-1) \ln\left(\sum_{m=1}^{M} |\lzero^{-1/2}\bm{\Delta}_m
|\right),
 \nonumber
 \end{eqnarray}
where we have dropped irrelevant constants arising from
normalisation. We now perform a procedure which is analogous to
the one used to obtain equation (\ref{fgauss}). In the polymer
melt before crosslinking, we assume that the ensemble of chains
obeys the distribution in (\ref{weight}) giving the free energy
per strand (\ref{free0}). The process of crosslinking not only
quenches the end points of each of the crosslinked strands, but
also quenches the nodes of the primitive path ${\bm\Delta}_m$,
since the crosslinked chains cannot disentangle due to the fixed
topology of the network. In our mean field approach, the tube
segments described by ${\bm\Delta}_m$ are conserved. For
evaluating the quenched average, note that the statistical weight
(\ref{weight}) treats all tube segments $m$ in an equivalent way.
This allows one to perform the summation over the index $m$,
separating the diagonal and the off-diagonal terms:
\begin{eqnarray}
    \beta F &=&\frac{3}{2b N}\left( M \langle
\bm{\Delta}_m^{\mathsf{T}}\cdot\lzero^{-1}\cdot\bm{\Delta}_m
\rangle \right.     \label{free} \\
&& \ \ \ \ \left. + M(M-1)\langle |\lzero^{-1/2}\bm{\Delta}_m |
|\lzero^{-1/2}\bm{\Delta}_n | \rangle \right)
    \nonumber\\
&&+ (M-1)\left\langle \ln\left(\sum_{m=1}^{M}
|\lzero^{-1/2}\bm{\Delta}_m | \right) \right\rangle, \nonumber
\end{eqnarray}
for arbitrary values of $m$ and $n\neq m$; the brackets
$\langle\cdots\rangle$ refer to the average with the probability
$P(\{\bm{\Delta}_m\})$ given in (\ref{weight}).

Any mechanical deformation expressed by the general strain tensor
$\matr{\lambda}$ will affinely transform $\bm{\Delta}_m$ into
$\bm{\Delta}_m^{'} =\matr{\lambda}\bm{\Delta}_m$. It could also
affect the nematic order: the director $\bm{n}$ could adopt a
different orientation under deformation and the degree of average
chain anisotropy $r$ may change as well. In other words, the
matrix $\lzero$, which characterises the anisotropy of the steps,
transforms into a new matrix $\ltheta$ with different eigenvalues
$l^{\parallel}$ and $l^{\perp}$ in a reference frame rotated by
the angle $\theta$. Hence $|\lzero^{-1/2} \bm{\Delta}_m|$
transforms into $|\ltheta^{-1/2} \matr{\lambda} \bm{\Delta}_m|$ on
deformation, but the distribution $P(\{\bm{\Delta}_m\})$ remains
unchanged. Bearing this in mind, we can evaluate the averages
(\ref{free}), leading to the free energy per crosslinked chain.
The Appendix gives a more detailed account of how one evaluates
the averages. The resulting elastic energy density takes the form
\begin{eqnarray}
    F_{\rm el}&=&\frac{2}{3}\mu\ \frac{2M+1}{3M+1} \,
    \mathrm{Tr}(\lzero\cdot\matr{\lambda}^{\mathsf{T}}
    \cdot\ltheta^{-1}\cdot\matr{\lambda})
        \label{aniso-free-energy} \\
    &+& \frac{3}{2}\mu(M-1)\frac{2M+1}{3M+1}
\left( \overline{
|\ltheta^{-1/2}\cdot\matr{\lambda}\cdot\lzero^{1/2}| } \right)^2\nonumber\\
    &+& \mu(M-1) \overline{ \ln|\ltheta^{-1/2}\cdot\matr{\lambda}\cdot\lzero^{1/2}|} ,
\nonumber
\end{eqnarray}
where we use the notations:
\begin{eqnarray}
\overline{ |\ltheta^{-1/2}\cdot\matr{\lambda}\cdot\lzero^{1/2}| }
&=& \frac{1}{4\pi}\!\int\limits_{|{\bf e}|=1}\!\!d\Omega
    |\ltheta^{-1/2}\cdot\matr{\lambda}\cdot\lzero^{1/2}\ {\bf e}|
        \label{nemmodmat}\hspace{10mm}\\
\overline{
\ln|\ltheta^{-1/2}\cdot\matr{\lambda}\cdot\lzero^{1/2}|} &=&
        \frac{1}{4\pi}\!\!\int\limits_{|{\bf e}|=1}\!\!d\Omega
    \ln|\ltheta^{-1/2}\cdot\matr{\lambda}\cdot\lzero^{1/2}\
     {\bf e}|\label{nemlogmat}
\end{eqnarray}
(the overline notation $\overline{\cdots }$ refers to the angular
averaging over the orientations of an arbitrary unit vector {\bf
e} used to contract a corresponding matrix into a vector, before
calculating its absolute value).

Expressions (\ref{nemmodmat}) and (\ref{nemlogmat}) can be
evaluated in various particular cases of deformation
$\matr{\lambda}$ and director orientation. The Appendix gives a
result for uniaxial deformation along the director, where
$\matr{\lambda}$ takes a diagonal form with
$\lambda^{\parallel}=\lambda$ and
$\lambda^{\perp}=1/\sqrt{\lambda}$. Explicit formulae for
(\ref{nemmodmat}) and (\ref{nemlogmat}) need to be inserted into
(\ref{aniso-free-energy}) to give the full elastic energy.

\section{Discussion}

From the expression (\ref{aniso-free-energy}), we can recover the
elastic free energy of an ideal phantom-chain nematic network by
taking the case $M=1$. This limit means physically that the
polymer strand is placed in one single tube, tightly confined to
the axis. Mathematically, a random walk in three dimensions with
$N$ steps is equivalent to a random walk in one dimension along a
given direction with $N/3$ steps. This fact is the underlying
reason why we recover the phantom chain network result by taking
$M=1$ in our model.

On the other hand, as the number of tube segments $M$ becomes
large, one obtains a rubber-elastic elastic energy of the form
 \beq
F_{\rm el}= \mu M\left(
(\langle|\ltheta^{-1/2}\cdot\matr{\lambda}\cdot
\lzero^{1/2}|\rangle)^2 +
\langle\ln|\ltheta^{-1/2}\cdot\matr{\lambda}\cdot
\lzero^{1/2}|\rangle \right).
  \label{dense-limit}
 \eeq
There are two ways to have a physical situation corresponding to
this limit of $M\gg 1$: either the polymer melt is very dense,
causing a high entanglement density, or the polymer chain is very
long between its crosslinked ends. In the latter case, the polymer
strand experiences many confining entanglements along its path.

Recall that the $F_{\rm el}$ is the elastic energy density, which
relates to the free energy per chain $F$ as: $F_{\rm el}=n_{\rm
ch} F$, where $n_{\rm ch}$ is the density of crosslinked strands.
We can assume that in a polymer melt, the chain density is
inversely proportional to the volume of an average chain, hence
inversely proportional to the contour length of this chain:
$n_{\mathrm{ch}}\propto 1/L$. In case of the phantom chain network
the rubber modulus $\mu=n_{\rm ch}k_B T$ [equation
(\ref{fgauss})]. One concludes in this case that the elastic
energy $F_{\mathrm{el}}$ scales with $1/L$, and therefore
$F_{\mathrm{el}}\rightarrow 0$ as the chains become infinitely
long! This unphysical behaviour reflects the fact that the phantom
chain model assumes the entanglement interactions of the chains
irrelevant. Clearly, this assumptions breaks down in the long
chain limit, where one expects the entanglements to play a crucial
role.

This unphysical behaviour is overcome by our expression
(\ref{dense-limit}). As the strands become longer, they will
experience more entanglements, generating more confining tube
segments. We could reasonably assume that the number of
entanglements and therefore the number of tube segments scales
linearly with the strand length $L$: $M\propto L$. Considering
expression (\ref{dense-limit}), one can note that the
corresponding rubber modulus does not vanish in the limit
$L\rightarrow\infty$, but remains a constant corresponding to the
``rubber plateau'' in a densely entangled melt.

Considering the particular case of uniaxial strain along the
constant nematic director $\bm{n}$, one can examine one of the key
physical effects found in nematic elastomers -- the spontaneous
mechanical deformations as the degree of anisotropy is changed,
for instance, by changing the temperature (and thus the nematic
order parameter $Q(T)$ and the effective chain anisotropy $r$).
Within the ideal phantom-chain model (\ref{ftrace}), applying a
uniaxial deformation along the director $\bm{n}$ with
$\lambda^{\parallel}=\lambda$ and $\lambda^{\perp}=
1/\sqrt{\lambda}$, one obtains
$$
    F_{\rm el}= \frac{1}{2}\mu\left(
    \frac{l_0^{\parallel}}{l^{\parallel}}\lambda^2+
    2\frac{l_0^{\perp}}{l^{\perp}}\frac{1}{\lambda}
    \right),
$$
where $l_0^{\parallel}$ and $l_0^{\perp}$ are the principal values
of  $\lzero$, and similarly $l^{\parallel}$ and $l^{\perp}$ the
ones of $\ltheta$, the anisotropy of a state after the deformation
(of course, in this case no director rotation occurs). The free
energy is minimised by the strain
$$\lambda_{\mathrm{m}}=
(l^{\parallel}l_0^{\perp}/ l_0^{\parallel}l^{\perp})^{1/3},$$
which describes a spontaneous uniaxial deformation of a nematic
rubber, first discovered theoretically in \cite{wargel88} and
mentioned in the literature ever since. For instance, if the
initial state \lzero is isotropic (at $T>T_{\rm ni}$), then
$\lambda_{\mathrm{m}}= (l^{\parallel}/l^{\perp})^{1/3}$, a
function of nematic order parameter $Q(T)$ and could reach a
remarkable value of 400\% uniaxial extension in a highly
anisotropic main-chain nematic rubber \cite{berfin97}.

This result is not altered by the complicated additional terms in
(\ref{aniso-free-energy}): remarkably, exactly the same
deformation $\lambda_{\mathrm{m}}$ minimises all three
corresponding expressions derived from (\ref{aniso-free-energy}),
which are given in the Appendix.

If we now assume that both the chain anisotropy and the director
$\bm{n}$ are kept fixed under the deformation, $\ltheta=\lzero$,
and that the strain tensor $\matr{\lambda}$ is diagonal in the
reference frame of the anisotropy matrix, then we observe that the
matrices in (\ref{aniso-free-energy})--(\ref{nemlogmat}) are all
diagonal. Hence the anisotropic terms cancel each other out, and
we are left with the same elastic energy as in the isotropic case
\cite{ouriso}. Hence, even if the material is anisotropic, its
linear elastic modulus does not depend on the orientation under
the above assumptions of unchanged degree of anisotropy
$r=l^{\parallel}/l^{\perp}$: the Young's moduli
$E^{\parallel}=E^{\perp}=E$. However, the modulus for simple shear
is partially affected by the anisotropy of the nematic rubber.
Consider $\lambda_{ij}=\delta_{ij}+\varepsilon u_i v_j$, with
${\bm u}$ and ${\bm v}$, the two orthogonal unit vectors defining
the simple shear. If the deformation does not mix the parallel and
perpendicular directions, i.e., if ${\bm u}$ and ${\bm v}$ are
both perpendicular to ${\bm n}$, then the shear modulus is the
same as in the isotropic case,
\begin{equation}
 G=\frac{1}{3}E=\mu\left(
        \frac{4}{3}\cdot\frac{2M+1}{3M+1}
+\frac{1}{5}(M-1) \cdot \frac{11M+5}{3M+1}  \right) . \nonumber
\end{equation}
On the other hand, if one of the vectors ${\bm u}$ or ${\bm v}$ is
parallel to the director ${\bm n}$, then the shear modulus is
changed by a factor of $(l_0^{\perp}/l_0^{\parallel})^{1/2}$ or
$(l_0^{\parallel}/l_0^{\perp})^{1/2}$, respectively.

If the material is not allowed to deform, any rotation of the
nematic director ${\bm n}$ away from its equilibrium orientation
${\bm n}_0$ will cost energy. In phantom-chain networks, the trace
formula (\ref{ftrace}) gives the corresponding elastic free energy
increase as a function of $\theta$, the angle between ${\bm n}$
and ${\bm n}_0$:
$$
    \Delta F_{\rm el}=\frac{1}{2}\mu
\left(\frac{l_0^{\perp}}{l_0^{\parallel}}+
\frac{l_0^{\parallel}}{l_0^{\perp}}-2\right)\sin^2 \theta
 \approx \frac{1}{2}\mu\frac{(r-1)^2}{r} \theta^2
$$
(in the limit of small director rotation $\theta$). This gives the
expression for the relative rotation coefficient $D_1$, first
written down phenomenologically by de Gennes \cite{gennes80} and
extensively discussed in the literature
\cite{warter96,brafin98,ter99}. In the small strain limit
$\matr{\lambda}=\underline{\underline{\delta}}
+\matr{\varepsilon}$, the coupling between the director rotation
${\bm \omega}=[{\bm n}\times\delta{\bm n}]$ and the antisymmetric
part of the strain $\Omega_i=\epsilon_{ijk} \varepsilon_{jk}$ can
be written as:
$$
\frac{1}{2}D_1\left[{\bm n}\times({\bm \Omega}-{\bm
\omega})\right]^2 +D_2 \, {\bm n}\cdot
\matr{\varepsilon}^{(S)}\cdot \left[{\bm n}\times({\bm
\Omega}-{\bm \omega})\right],
$$
where $\underline{\underline{\varepsilon}}^{(S)}$ is the symmetric
part of the small strain. The entanglement model does not change
the dependence $D(r)$ qualitatively, but introduces a coefficient
associated with the entanglement density:
\begin{eqnarray*}
   D_1 = \mu\frac{(r-1)^2}{r}
        \left(\frac{33M^2+22M +5}{30(3M+1)}\right) \
   \approx 0.4 \mu \, M \, \frac{(r-1)^2}{r}.
\end{eqnarray*}

Another key physical property of nematic rubbers is the effect of
soft elasticity. Fundamental internal symmetries of an elastic
medium with an independently mobile orientational degree of
freedom, the nematic director $\bm{n}$, demand that there is a
particular relationship between the two relative rotation
coefficients $D_1$ and $D_2$ and one of the linear shear moduli,
$C_5$, \cite{gollub}. It has been shown \cite{olmsted} that there
is a continuous set of such soft deformations (not necessarily
small in amplitude), which by appropriately combining strains and
director rotations can make the elastic response vanish
completely:
$$\matr{\lambda}_{\rm soft}=\ltheta^{1/2} \cdot \matr{U} \cdot
\lzero^{-1/2},$$
 where $\matr{U}$ is an arbitrary unitary (3D rotation) matrix.
It is quite obvious that substituting this strain tensor into the
modified tube-model expression (\ref{aniso-free-energy}) will
leave this free energy at its ground state level as well. It is,
in fact, gratifying that these two crucial physical effects
(thermal expansion and soft elasticity), which have attracted so
much theoretical and experimental attention in recent years, are
left intact within a much more complex theoretical description of
a highly entangled nematic elastomer.

\subsection*{Conclusion}

In this present work, we have analysed the behaviour of a uniaxial
nematic polymer network in the presence of chain entanglements,
which are treated within a tube model approach. We found that this
leads to a significantly modified rubber-elastic energy which, in
principle, should supersede the earlier molecular theory
(\ref{ftrace}). The present model captures the physics of
entanglements in a consistent way and, for the first time, takes
into account an orientational effect of chain conformation in the
tube segments aligned at an arbitrary angle with respect to the
uniform nematic director $\bm{n}$. Since the role of entanglements
is, from all points of view, much more significant in a
crosslinked network, the theory provides a firmer ground for
description of many theoretically known and experimentally tested
results.

We have to remark that our model only describes the equilibrium
response of a network to deformation. Shortly after applying the
deformation, the network will need to find a new microscopic
equilibrium. Each polymer strand would redistribute the monomers
between the affinely modified tube segments, attributing more
monomers to some segments, less to others, and eventually reaching
a new optimal conformation $\{ \overline{s}_m \}$. This gives the
expression for the rubber elastic free energy density
(\ref{aniso-free-energy}). The dynamics of this relaxation is
based on the sliding (reptation) motion along the primitive path
while constraining the end points of it. This process would be
reflected in a time dependence of the variable $s_{m}$, which is
the number of monomer steps attributed to the tube segment $m$. By
describing this relaxation process, one could extend the present
equilibrium model to describe the stress relaxation and the short
time viscoelastic response of a nematic
rubber. \\

We appreciate many useful discussions with S.F. Edwards and M.
Warner. S.K. gratefully acknowledges support from an Overseas
Research Scholarship, from the Cambridge Overseas Trust and from
Corpus Christi College.

\appendix

\section{Evaluation of quenched averages}

To evaluate averages $\langle
\bm{\Delta}_m^{\mathsf{T}}\cdot\lzero^{-1}\cdot\bm{\Delta}_m
\rangle$, $\langle |\lzero^{-1/2}\bm{\Delta}_m |
|\lzero^{-1/2}\bm{\Delta}_n | \rangle $ and $\langle\ln(\sum
|\lzero^{-1/2}\bm{\Delta}_m |)\rangle$ in equation (\ref{free}),
for the arbitrary $m,n=1,\dots,N$, one needs to integrate the
corresponding scalar functions of $\bm{\Delta}_m$ with respect to
the probability distribution (\ref{weight}). For this purpose, one
has first to find the normalisation $\hat{\cal N}$ of the
distribution, which can most easily be achieved by introducing a
new scalar variable $u=\sum_{m=1}^M |\lzero^{-1/2}\bm{\Delta}_m |
$ to simplify the exponent. It is also useful to change the
integration variables from $\bm{\Delta}_m$ to a transformed vector
$\widetilde{\bm{\Delta}}_m = \lzero^{-1/2}\bm{\Delta}_m$. One
obtains then:
\begin{eqnarray*}
    \hat{\cal N}&=&
     \prod_{m=1}^{M}\int d\bm{\Delta}_m
      \frac{\exp\left(-\frac{3}{2b^2N}
   \left(\sum_{m=1}^{M} |\lzero^{-1/2}\bm{\Delta}_m | \right)^2\right)
  }{\left(\sum |\lzero^{-1/2}\bm{\Delta}_m | \right)^{M-1}} \nonumber\\
    &=& \prod_{m=1}^{M}\int d {\bm{\Delta}}_m
\int\limits_0^{\infty} \frac{du}{u^{M-1}}\,
e^{-\frac{3}{2b^2N}u^2} \delta\left(u-\sum_{m=1}^M
|\lzero^{-1/2} {\bm{\Delta}}_m |\right) \nonumber\\
    &=&(4\pi)^{M} {\rm Det}\lzero^{M/2} \int\limits_0^{\infty}
    \frac{du}{u^{M-1}}\, e^{-\frac{3}{2b^2N}u^2}  \nonumber \\
    && \qquad \qquad \cdot \underline{
    \int\limits_0^u d\widetilde{\Delta}_1 \widetilde{\Delta}_1^2
        \int\limits_0^{u-|\widetilde{\bm{\Delta}}_1|}\hspace{-2mm}
        d\widetilde{\Delta}_2 \widetilde{\Delta}_2^2 \cdots } \nonumber\\
    &&  \cdots \underline{\int\limits_0^{u-\dots-|\widetilde{\bm{\Delta}}_{M-2}|}\hspace{-5mm}
        d\widetilde{\Delta}_{M-1} \widetilde{\Delta}_{M-1}^2\
        \left(u-{\textstyle{\sum_{m=1}^{M-1}}} |\widetilde{\bm{\Delta}}_m |
        \right)^2}
\end{eqnarray*}

In the last step, we introduced spherical coordinates for the
variables $\widetilde{\bm{\Delta}}_m$, implemented the
delta-function constraint $u=\sum \widetilde{\Delta}_m$ and used
the fact that the variables $\widetilde{\Delta}_m$ are bound to be
positive. The underlined expression is a multiple integral over
the hyper-triangular domain in the space of
$\{\widetilde{\bm{\Delta}}_m \}$ and is a function of $u$, which
we call $I_M(u)$. Since the integrals only involve power
functions, $I_M(u)$ itself is a power in $u$. It is then evaluated
via the iterative procedure, which generates the recursive
relation and returns an explicit function:
$$I_M=\frac{u^{3M-1}}{f_M} , \ \ {\rm with} \
f_M= \prod_{m=1}^{M-1} \frac{3m(3m+1)(3m+2)}{2}.$$
 The first two terms in the Eq.~(\ref{free}) involve the diagonal
($\sim \widetilde{\bm{\Delta}}_m^2$) or the off-diagonal ($\sim
\widetilde{\bm{\Delta}}_m \widetilde{\bm{\Delta}}_n$) factors. In
both cases, the integration procedure is analogous to that of
normalisation factor $\hat{\cal N}$ above, except that either one
($m$) or two ($m\neq n$) integrals in the sequence contain an
extra scalar factor of $\widetilde{\Delta}_m$. The corresponding
angular integration over the orientations of
$\widetilde{\bm{\Delta}}_m$ (producing a factor of $4\pi$ in
$\hat{\cal N}$) now becomes non-trivial, depending on its angle
relative to tensors $\matr{\lambda}$ and $\ltheta$, when the
sample is deformed. This angular integration is left unfinished
here, since it depends on particular deformation and director
geometry; the main thermodynamic average of the diagonal (square)
term returns the ideal trace-formula in the final free energy
density (\ref{aniso-free-energy}), while the off-diagonal average
returns the expression (\ref{nemmodmat}).

For the logarithmic term in (\ref{aniso-free-energy}), one
obtains:
\begin{eqnarray*}
\lefteqn{\left\langle\ln\left( \sum_{m=1}^M |\ltheta^{-1/2}\cdot
\matr{\lambda} \bm{\Delta}_m | \right)\right\rangle } \nonumber\\
    &=& \frac{{\rm Det}\lzero^{M/2}}{\hat{\cal N}}
\int_0^{\infty} \frac{du}{u^{M-1}}\, e^{-\frac{3}{2b^2N}u^2}
  \nonumber \\
 &&  \cdot \int_0^u d\widetilde{\Delta}_1 \widetilde{\Delta}_1^2
  \int_0^{u-|\widetilde{\bm{\Delta}}_1|}\hspace{-2mm} d\widetilde{\Delta}_2
  \widetilde{\Delta}_2^2 \cdots \nonumber\\
&& \cdots \int_0^{u-\dots
-|\widetilde{\bm{\Delta}}_{M-2}|}\hspace{-5mm}
 d\widetilde{\Delta}_{M-1} \widetilde{\Delta}_{M-1}^2\
 (u-{\textstyle{\sum_{m=1}^{M-1}}} |\widetilde{\bm{\Delta}}_m |)^2 \\
&& \cdot \left(\prod_{m=1}^{M}\int d\widetilde{\Omega}_m\right)
\ln \bigg[ \widetilde{\Delta}_1|\ltheta^{-1/2} \cdot
\matr{\lambda}\cdot
\lzero^{1/2} \widetilde{\bf e}_1| +\dots  \\
&& \cdots + \widetilde{\Delta}_{M-1}|\ltheta^{-1/2} \cdot
\matr{\lambda} \cdot \lzero^{1/2} \widetilde{\bf e}_{M-1}|          \nonumber\\
&& + \left(u-{\textstyle{\sum_{m=1}^{M-1}}}
|\widetilde{\bm{\Delta}}_m | \right)
    |\ltheta^{-1/2} \cdot \matr{\lambda} \cdot \lzero^{1/2} \widetilde{\bf e}_M| \bigg]
\end{eqnarray*}
Here $d\widetilde{\Omega}_m$ is the angular measure of
orientations of the corresponding unit vector $\widetilde{\bf
e}_m$, along the modified tube segment vector
$\widetilde{\bm{\Delta}}_m$. In the last term under the logarithm,
the absolute value of $\widetilde{\bm{\Delta}}_M$ is substituted
by its value from the delta-function constraint. The next step is
to approximate the logarithm with its complicated
angular-dependent argument:
\begin{eqnarray*}
\lefteqn{\ln[ \widetilde{\Delta}_1|\ltheta^{-1/2} \cdot
\matr{\lambda} \cdot \lzero^{1/2} \widetilde{\bf e}_1|+\dots}   \nonumber\\
    && \qquad +\Delta_{M-1}|\ltheta^{-1/2} \cdot \matr{\lambda} \cdot
\lzero^{1/2} \widetilde{\bf e}_{M-1}| \nonumber \\
    && \qquad +(u-\sum |\widetilde{\bm{\Delta}}_m |)
|\ltheta^{-1/2} \cdot \matr{\lambda} \cdot \lzero^{1/2} \widetilde{\bf e}_M|]\nonumber\\
    &=&\ln[u]+\ln |\ltheta^{-1/2} \cdot \matr{\lambda} \cdot
    \lzero^{1/2} \widetilde{\bf e}_M| + \ln \bigg[ 1+ \\
    && + \underbrace{\sum_{m=1}^{M-1}
        \left( \frac{|\ltheta^{-1/2} \cdot
\matr{\lambda} \cdot \lzero^{1/2} \widetilde{\bf e}_m|}
        {|\ltheta^{-1/2} \cdot \matr{\lambda} \cdot \lzero^{1/2} \widetilde{\bf
        e}_M|}-1 \right)
\frac{\widetilde{\Delta}_m}{u}}_{\mbox{small since }u\gg
\widetilde{\Delta}_m}\bigg] \nonumber\\
&& \approx \ln |\ltheta^{-1/2} \cdot \matr{\lambda} \cdot
\lzero^{1/2} \widetilde{\bf e}_M|+\mathrm{const}.
\end{eqnarray*}
After this, all of the results of multiple integrals over
$d\widetilde{\Delta}_m$ and $d\widetilde{\Omega}_m$ cancel with
the normalisation factor and the only relevant contribution arises
from the angular integration of the scalar logarithmic term over
the orientations of ${\bf e}_M$, cf. equation (\ref{nemlogmat}).

In the particular case of uniaxial deformation along the nematic
director $\bm{n}_0$, with $\lambda^{\parallel}=\lambda$ (along
$\bm{n}_0$=const) and $\lambda^{\perp}=1/\sqrt{\lambda}$, the
evaluation of equations
(\ref{aniso-free-energy})-(\ref{nemlogmat}) gives:
\begin{eqnarray*}
    \mathrm{Tr}(\lzero\cdot\matr{\lambda}^{\mathsf{T}}\cdot\ltheta^{-1}\cdot\matr{\lambda})
    &=&
    \left( \frac{l_0^{\parallel}}{l^{\parallel}} \right) \lambda^2+
    2\left( \frac{l_0^{\perp}}{l^{\perp}} \right) \frac{1}{\lambda} \nonumber \\
\overline{ |\ltheta^{-1/2}\cdot\matr{\lambda}\cdot\lzero^{1/2}| }
    &=&\frac{1}{2}\Bigg(
        \sqrt{\frac{l_0^{\parallel}}{l^{\parallel}}}\lambda
        +\sqrt{r} \sqrt{\frac{l_0^{\perp}}{l^{\perp}}}
        \frac{\ln\left(\frac{\lambda^{3/2}+\xi}{\lambda^{3/2}-\xi}\right)}
{2\sqrt{\lambda}\xi}  \Bigg)   \nonumber  \\
\overline{ \ln|\ltheta^{-1/2}\cdot\matr{\lambda}\cdot\lzero^{1/2}|
} &=& \ln(\lambda)-1\nonumber\\
    &&+\sqrt{r}\frac{\arctan (\xi / \sqrt{r}) }{\xi}
    +\frac{1}{2}\ln\left(\frac{l_0^{\parallel}}{l^{\parallel}}\right)
\end{eqnarray*}
with the notations
$$
    \xi=\sqrt{\lambda^3-r} \qquad \mbox{and} \qquad
    r=\frac{l_0^{\perp}l^{\parallel}}{l^{\perp}l_0^{\parallel}}.
$$
If we assume that the anisotropy is not changed by the
deformation, i.e. $\lzero=\ltheta$, (and the director preserves
its original orientation $\bm{n}_0$) then the elastic response of
the nematic rubber is not different from isotropic behaviour
\cite{ouriso}.


\begin{thebibliography}{22}

\bibitem{gennes79}
P.~G. de Gennes, {\em Scaling Concepts in Polymer Physics}
(Cornell University Press, Ithaca, N.Y., 1979).

\bibitem{doiedw86}
M. Doi and S.~F. Edwards, {\em Theory of Polymer Dynamics}
(Clarendon Press, Oxford, 1986).

\bibitem{jarry-monnerie79}
J.-P. Jarry and L. Monnerie, Macromolecules {\bf 12},  316  (1979).

\bibitem{deloche-samulski81}
B. Deloche and E.~T. Samulski, Macromolecules {\bf 14},  575  (1981).

\bibitem{doi-pearson89}
M. Doi, D. Pearson, J. Kornfield, and G. Fuller, Macromolecules {\bf 22},  1488
   (1989).

\bibitem{bladon-warner93}
P. Bladon and M. Warner, Macromolecules {\bf 26},  1078  (1993).

\bibitem{abramchuk-nyrkova89}
S.~S. Abramchuk, I.~A. Nyrkova, and A.~R. Khokhlov, Polymer
Science U.S.S.R.
  {\bf 31},  1936  (1989).

\bibitem{warter96}
M. Warner and E.~M. Terentjev, Prog. Polym. Sci., {\bf 21}, 853
(1996).

\bibitem{brafin98}
H.~R. Brand and H. Finkelmann, in: {\it Handbook of Liquid
Crystals}, ed D. Demus et al. (Wiley-VCH, Weinheim, 1998), Vol.3,
Chapter V.

\bibitem{ter99}
E.~M. Terentjev, J. Phys. Cond. Mat., {\bf 11}, R239 (1999).

\bibitem{edward77}
S.~F. Edwards, Brit. Polymer J.  140  (1977).

\bibitem{gaydou90}
R.~J. Gaylord and J.~F. Douglas, Polymer Bulletin {\bf 23},  529
(1990).

\bibitem{baldoi81}
R.~C. Ball, M. Doi, S.~F. Edwards, and M. Warner, Polymer {\bf
22},  1010
  (1981).

\bibitem{higbal89}
P.~G. Higgs and R.~C. Ball, Europhysics Letters {\bf 8},  357
(1989).

\bibitem{ouriso}S. Kutter and E.~M. Terentjev --
submitted (2001), cond-mat/0106371.

\bibitem{wang-warner86}
X.~J. Wang and M. Warner, J. Phys. A {\bf 19},  2215  (1986).

\bibitem{wargel88}
M. Warner, K.~P. Gelling, and T.~A. Vilgis, J. Chem. Phys. {\bf
88},  4008 (1988).

\bibitem{viledw}
S.~F. Edwards and T.~A. Vilgis, Rer. Prog. Phys. {\bf 51}, 243
(1988).

\bibitem{berfin97}
G.~H.~F. Bergmann, H. Finkelmann, V. Percec, and M. Zhao,
Macromol. Rapid. Commun. {\bf 18},  353  (1997).

\bibitem{gennes80}
P. G. de Gennes, in: {\it Liquid Crystals of One- and
Two-Dimensional  Order}, ed W. Helfrich and G. Heppke (Springer,
Berlin, 1980), p. 231.

\bibitem{gollub}L. Golubovi\'c and T.~C. Lubensky, Phys. Rev. Lett.
{\bf 63}, 1082 (1989).

\bibitem{olmsted}P.~D. Olmsted, J. Physique II {\bf 4} 2215 (1994).

\bibitem{higgay90}
P.~G. Higgs and R.~J. Gaylord, Polymer {\bf 31},  70  (1990).

\end{thebibliography}
\end{document}